\documentclass[12 pt]{article}
\usepackage[utf8]{inputenc}
\usepackage{authblk}
\usepackage[square,sort,comma,numbers]{natbib}
\usepackage{epigraph}
\usepackage{hyperref}
\usepackage{graphicx}
\usepackage{mathtools}   
\usepackage{amsfonts}
\usepackage{appendix}
\usepackage{caption}
\usepackage{orcidlink}
\hypersetup{
    colorlinks=true,
    linkcolor=blue,
}
\providecommand{\keywords}[1]
{
  \small	
  \textbf{\textit{Keywords---}} #1
}

\begin{document}

\title{Reassessing the problem of time of quantum gravity}

\author[1,2]{Álvaro Mozota Frauca}
\affil[ ]{alvaro.mozota@uab.cat, \orcidlink{0000-0002-7715-0563} \href{https://orcid.org/0000-0002-7715-0563}{https://orcid.org/
0000-0002-7715-0563}}
\affil[1]{Department of Philosophy. Universitat Aut\`onoma de
Barcelona, Building B Campus UAB 08193 Bellaterra (Barcelona), Spain}
\affil[2]{LOGOS, University of Barcelona, Department of Logic, History and
Philosophy of Science, Carrer de Montalegre 6 08001 Barcelona, Spain}
\date{\today}

\maketitle

%
%
%



\begin{abstract}
In this paper I raise a worry about the most extended resolutions of the problem of time of canonical quantizations of general relativity. The reason for this is that these resolutions are based on analogies with deparametrizable models for which the problem can be solved, while I argue in this paper that there are good reasons for doubting about these resolutions when the theory is not deparametrizable, which is the case of general relativity. I introduce an example of a non-deparametrizable model, a double harmonic oscillator system expressed by its Jacobi action, and argue that the problem of time for this model is not solvable, in the sense that its canonical quantization doesn't lead to the quantum theory of two harmonic oscillators and the standard resolutions of the problem of time don't work for this case. I argue that as general relativity is strongly analogous to this model, one should take seriously the view that the canonical quantization of general relativity doesn't lead to a meaningful quantum theory. Finally, I comment that this has an impact on the foundations of different approaches to quantum gravity.
\end{abstract}




It is a well-known fact that the canonical quantization of general relativity, in any of its formulations, leads to a problem of time. When one applies the standard quantization rules for gauge theories to general relativity what one finds is a series of constraint equations but no dynamical equation which describes an evolution in time of the quantum state. There is a number of proposals for how to interpret this `timeless' quantum state and for how to recover the usual time evolution of our physical theories. However, most of these proposals are based on analogies with simpler models, which in this paper I will argue that are misleading. The reason for this is that most of these models correspond to deparametrizable theories, that is, theories in which time is represented as one of the variables of the theory, while general relativity is not deparametrizable, as the way temporal information is encoded in the theory is more sophisticated.

I will start by briefly reviewing the problem of time in section \ref{Problem_of_time}. First I will introduce the constrained formalism, an extension of the canonical formalism to be able to accommodate theories like gauge theories or reparametrization invariant theories. Next, I will sketch how the quantization procedure for these theories works: basically, the quantization procedure is the same as for any other theory expressed in the canonical formalism, but with the complication that we now have to impose a series of constraints to our quantum states. However, I will show how the imposition of the constraints for reparametrization invariant theories implies that there is no non-trivial quantum dynamics for these theories, which is known as the problem of time\footnote{Authors like Kuchar and Isham speak about the problems of time in the plural, as there are several technical and conceptual problems related to this.}.

In section \ref{deparametrizable_models} I will introduce a simple example: the case of a non-relativistic particle expressed in a reparametrization invariant way. I will show that the problem of time in this case can be solved and I will argue that the reason why this is so is that time is part of the configuration space of this model. In other words, we are able to identify a variable as time and deparametrize the theory, i.e., to express the dynamics not with respect to an arbitrary parameter but with respect to the physical time. I will further argue that this model of resolution of the problem of time can be extended, with some caveats, to other models, which are also deparametrizable. A model is deparametrizable if time, or spacetime coordinates, appear explicitly as configuration or phase space variables or can be identified and separated in some of these spaces by means of some coordinate transformation or canonical transformation. I will also argue that most of the attempts of resolution of the problem of time for the case of quantum gravity are based on cases similar to this one.

On the other hand, in section \ref{non-deparametrizable_models} I will introduce a different kind of models which also show a reparametrization invariance but which are not deparametrizable, that is, even if we can choose freely the way we parametrize our theory, this doesn't mean that time is encoded in the configuration space of the models. In particular, I will introduce a simple example of a model describing a system of two harmonic oscillators, and I will show how the resolutions of the problem of time for deparametrizable models fail in this case. In this sense, I argue that the problem of time is more serious for theories of this kind, and that it may even be unsolvable.

The relevant question to address if we are interested in the quantization of gravity is therefore whether general relativity is a deparametrizable theory or not. In section \ref{gr_not_deparametrizable} I briefly introduce general relativity in its canonical formulation and comment on some results which show that it is not deparametrizable. In this sense, in the same way that the problem of time wasn't solvable for the case of the two harmonic oscillators, I will argue that we may be in the same situation for the case of general relativity, and that the attempts of resolution may be misguided. I will also comment on some positions like Kiefer's \citep{Kiefer2012} which argue that even if general relativity is not deparametrizable, there is a sense in which spacetime is encoded in the configuration space of the theory and that some of the resolutions which applied to deparametrizable models may still apply to general relativity. I will argue that the way time is supposed to be encoded in the configuration space of general relativity is unclear and that the two arguments usually formulated to support this claim, namely, the thick sandwich conjecture and the counting of degrees of freedom argument, aren't valid, as they could also be applied to the double harmonic oscillator example to reach the wrong conclusion that time is encoded in the configuration space of the system. In this sense, the analogy between general relativity and my example is strong, which leads to the conclusion that the problem of time is serious in quantum gravity, perhaps even unsolvable.  


Finally, in section \ref{consequences} I comment on the impact that this analysis has on theories and models of quantum gravity. In particular, I will argue that the worry raised in this paper has an impact not only on quantum geometrodynamics, but also on the more modern LQG, on related approaches and also on cosmological models. My argument in this paper shows that all these theories and models are based on some foundations which are at least questionable.

\section{The problem of time} \label{Problem_of_time}

In this first section I will briefly review the way the problem of time arises for reparametrization invariant theories. First, I will introduce the constrained formalism which is used for describing the dynamics of singular systems like gauge systems or reparametrization invariant systems. Then, I explain how the presence of constraints alters the usual canonical quantization procedure. Finally, I introduce reparametrization invariant models and how the quantization procedure makes the quantum dynamics trivial for these models, which is known as the problem of time. I refer the reader to \cite{Sundermeyer1980, Henneaux1994,Rothe2010} for detailed introductions to the constrained formalism and its quantization. 

A word on notation. Throughout the paper I will be using Einstein's summation convention and every time there are repeated indices a sum over them will be assumed. I will be using greek indices $\mu, \nu$ to represent spacetime indices and latin indices $a, b$ to represent spacetime indices restricted to just the spatial ones. Other indices like $\alpha, A, i$ will be used for indexing constraints, fields or variables, and their meaning should be clear from the context.

\subsection{Constrained Hamiltonian dynamics}

The kind of models we are interested in are formally indeterministic, i.e., given a set of initial conditions at a time the equations of motion don't determine the final state at a later time. For instance, in the 4-potential version of electromagnetism, the gauge invariance of the theory makes it the case that an initial state determines the final 4-potential only up to a gauge transformation. Similarly, general relativity fails to determine uniquely the value of the metric at a coordinate point $x^{\mu}$, given that the diffeomorphism invariance of the theory makes it the case that the coordinates are meaningless, and there are different solutions of Einstein equations which assign different values of the metric to the same coordinate point. Notice that in both cases this indeterminism is just formal and that the theories are deterministic from the physical point of view: an initial configuration of the electromagnetic field determines it at any posterior (and previous) time and any configuration of the metric at a Cauchy slice determines the geometry of the whole spacetime. The formal indeterminism of these theories can be dealt with in the Hamiltonian formalism, which will be the starting point for the quantization of these theories. 

The first step for expressing a gauge theory or a reparametrization invariant theory in the Hamiltonian formalism is to define an action:
\begin{equation}
S[q_i]=\int dt L(q_i,\dot{q}_i,t) \, ,
\end{equation}
where $L$ is the Lagrangian of the theory, $q_i$ represent the different variables or fields in the theory. Imposing that physical trajectories minimize the action leads to the Euler-Lagrange equations, which contain the dynamics of the theory:
\begin{equation}
\frac{\delta S}{\delta q_i}=0 \rightarrow \ddot{q}_j\frac{\partial ^2 L}{\partial \dot{q}_j\partial \dot{q}_i}=\frac{\partial L}{\partial q_i} -\dot{q}_j\frac{\partial ^2 L}{\partial q_j   \partial \dot{q}_i} \, .
\end{equation}
This system of differential equations has a unique solution just in case one can invert the Hessian matrix $\frac{\partial ^2 L}{\partial \dot{q}_j\partial \dot{q}_i}$ and express the accelerations $\ddot{q}_i$ in terms of the positions and velocities. The theories we are interested in have multiple solutions, corresponding to different gauges or different parametrizations, and hence it is a necessary property of these theories that the Hessian matrix is not invertible. In this case we say that the Lagrangian is singular.

Let me mention that singular Lagrangians allow not only for indeterminism, but also it can be the case that the set of possible initial conditions is restricted, in the sense that there will be initial conditions for which there doesn't exist any solution of the Euler-Lagrange equations. For instance, this is the case for electromagnetism: only initial conditions satisfying Gauss law can satisfy the equations of motion. In the case of general relativity we will also find in section \ref{gr_not_deparametrizable} that not every initial condition is allowed. 

When one tries to express a theory with a singular Lagrangian in the Hamiltonian formulation, one finds the following problem. The canonical momenta $p^i$ are defined by the transformation $p^i=\frac{\partial L}{\partial \dot{q}_i}$, and this can be shown to be invertible if and only if the Hessian matrix is. As we are dealing with singular Lagrangians, the definition of the momenta is not invertible, and only a subregion of phase space corresponds to the image of $q_i, \dot{q}_i$ under this transformation. This subregion, the constraint surface, is defined as the region where a number of functions $\phi _{\alpha}$, known as primary constraints, vanish. Therefore, any physical evolution can be described by an evolution in the constraint surface and not as an evolution in the whole phase space.

There are different ways of defining the dynamics in this formalism. Here it will be enough for us to introduce the total Hamiltonian:
\begin{equation}
H_T=H_c+v^{\alpha}\phi_{\alpha} \, ,
\end{equation}
where $H_c$ is the canonical Hamiltonian defined by $H_c=\dot{q}_ip^i-L$ on the constraint surface but extended\footnote{As a function on the whole phase space there are different canonical Hamiltonians one can define. In any case, the dynamics they define are equivalent, corresponding to redefinitions of the free functions $v^{\alpha}$.} to be a function for the whole phase space and $v^{\alpha}$ are arbitrary functions. The dynamics for the phase space variables and any phase space function $f$ is defined by the Hamilton equations of motion:
\begin{eqnarray}
\dot{q}_i=\{q_i,H_T \}=\frac{\partial H_c}{\partial q_i} +v^{\alpha}\frac{\partial \phi_{\alpha}}{\partial q_i} \\
\dot{p}_i=\{p_i,H_T \}=-\frac{\partial H_c}{\partial p_i} -v^{\alpha}\frac{\partial \phi_{\alpha}}{\partial p_i} \\
\dot{f}(q,p,t)=\{f(q,p,t),H_T \}+\frac{\partial f(q,p,t)}{\partial t} \, ,
\end{eqnarray}
where it is assumed that the constraints are satisfied\footnote{Let me mention that in this paper and in the literature with the term `constraints' one refers both to certain functions $\phi ^{\alpha}$ or operators $\hat{\phi} ^{\alpha}$ and to the condition that these functions vanish or that the action of those operators on certain states vanishes.}, i.e., $\phi^{\alpha}=0$, and the brackets represent the Poisson brackets. In order for these dynamics to be consistent with the primary constraints it may be the case that further constraints need to be imposed, which are known as secondary constraints, and/or that the $v^{\alpha}$ are not arbitrary functions but are fixed. This latter case is uninteresting for the purposes of this paper, as we are interested in situations in which there is no condition on the $v^{\alpha}$ which makes it the case that the solutions of the equations of motion depend on arbitrary functions, showing the formal indeterminism of the theories we are studying\footnote{For completeness, let me say that systems like the ones we are interested in are called first-class systems, systems in which the $v^{\alpha}$ are fixed are second-class, and that there are systems in which one has a mixture of first-class and second-class constraints}.

Finally, in the constrained formalism one can define gauge generators, which generate gauge transformations which transform solutions of the equations of motion into other solutions of the equation of motion with the same physical content. Gauge generators take the form of a linear combination of constraints\footnote{In the literature there is some confusion about gauge transformations and their generators. It is sometimes claimed that first-class constraints generate gauge transformations, but as it is argued by Pitts \cite{Pitts2013} this claim is not true, and the generators take the form of some specific linear combination of the constraints.}, multiplied by a number of functions $\epsilon$. In this sense, any phase space function $f$ under an infinitesimal gauge transformation is changed by:
\begin{equation}
\delta f =\{f,G[\epsilon]\} \, .
\end{equation}
For gauge theories like electromagnetism one can identify the gauge-invariant quantities as the quantities which don't change under a gauge transformation, that is, the ones which have vanishing Poisson brackets with the gauge generator. This way of defining gauge transformations and gauge invariance is a useful characterization in the constrained formalism, although we will see that it may be misleading in the case of reparametrization invariant theories. Now we can turn to study the quantization of constrained systems.


\subsection{Quantization}

Let me start by reviewing the canonical quantization procedure for an unconstrained system. At a kinematical level, the quantization requires defining a Hilbert space in which some of the functions of the phase space of the theory are represented as linear operators. The algebra of operators intends to mimic the Poisson algebra of the classical theory:
\begin{equation} \label{correspondence_Poisson_commutators}
[\hat{f},\hat{g}]=i\hbar \{ \widehat{f,g} \}\, .
\end{equation}
However, the quantization cannot be complete in the sense that there is no way of assigning an operator to every phase space function such that for any two arbitrary functions \ref{correspondence_Poisson_commutators} is satisfied\footnote{This is a well-established fact proved as a theorem in \cite{Groenewold1946}.}. Therefore, one chooses a restricted algebra of functions and quantizes it, while for the rest of phase space functions there will remain some ambiguity. For instance, for the quantum mechanics of a single particle one quantizes $q$ and $p$ so that their associated operators satisfy $[\hat{q},\hat{p}]=i\hbar$. In this case there is some ambiguity in defining a quantum operator for the function $qp$, as both (but not only) $\hat{q}\hat{p}$ and $\hat{p}\hat{q}$ could be the quantum counterpart of this function.

For having a complete quantum theory one needs to define a Hamiltonian operator, which will define the dynamics of the theory. In general, there will be some ambiguity in quantizing the classical Hamiltonian function, and different choices will be available, giving rise to different quantum theories. Once a choice is made, it defines the dynamics for the expectation value of any operator by means of:
\begin{equation}\label{quantum_dynamics_canonical}
\frac{d}{dt}\langle \hat{O} \rangle =\frac{1}{i\hbar} \langle [\hat{O}, \hat{H} ] \rangle + \langle \frac{\partial \hat{O}}{\partial t}  \rangle \, .
\end{equation}
This form of expressing the dynamics is equivalent to more familiar representations such as the Schr\"odinger or the Heisenberg pictures.
 
I summarize the correspondence between the classical and the quantum theory in table \ref{Table_Quantum_Classical_correspondence} and steps of the quantization process as:
\begin{enumerate}
\item{Start with a classical theory defined on a phase space.}
\item{Choose a subalgebra of functions on phase space and quantize them, i.e., build an algebra of operators on a Hilbert space $\mathcal{H}$ such that the commutator algebra is defined by the Poisson algebra of the classical functions (eq. \ref{correspondence_Poisson_commutators}).}
\item{Build a Hamiltonian operator which is a quantization of the classical one. The dynamics of the theory is contained in equation \ref{quantum_dynamics_canonical} or in some equivalent form.}
\end{enumerate}

\begin{table}[ht]
\centering

\begin{tabular}[t]{lccl}
\hline
 & Classical Theory & Quantum Theory                               \\ \hline
Basic space                       & Phase space $\mathcal{P}$                           & Hilbert space $\mathcal{H}$                                    \\ 
Observables            & Functions on     $\mathcal{P}$                       & Operators on $\mathcal{H}$                            \\
Algebra             & Poisson algebra                         & Commutator algebra                           \\ 
Dynamics             & $\dot{f}(q,p,t)=\{f(q,p,t),H \}+\frac{\partial f(q,p,t)}{\partial t}$                         & $\frac{d}{dt}\langle \hat{O} \rangle =\frac{1}{i\hbar} \langle [\hat{O}, \hat{H} ] \rangle + \langle \frac{\partial \hat{O}}{\partial t}  \rangle$                           \\
\hline
\end{tabular}
\caption{Correspondence between the elements of a classical and a quantum theory.}
\label{Table_Quantum_Classical_correspondence}
\end{table}

This quantization procedure can be adapted to constrained systems\footnote{This is usually known as Dirac quantization, as it was first formulated by Dirac in \cite{Dirac1964}. I refer the reader again to \cite{Sundermeyer1980, Henneaux1994,Rothe2010} for more careful discussions.}. In the canonical description of a constrained system, not every point in phase space represents physically meaningful states, as these are restricted to the constraint surface. Similarly, in the quantum description we will distinguish between two Hilbert spaces: a `bigger' one which is the counterpart of the whole phase space and a `smaller' one which will be the counterpart of the constraint surface. The bigger space is known as the kinematical Hilbert space and it is defined in the same way as the Hilbert space of an unconstrained system, i.e., it is a Hilbert space in which a series of operators exists such that their operator algebra mimics an algebra of functions in phase space. The physical Hilbert space is defined as the subspace of this space, or as the distributional space over this space for infinite-dimensional systems, of states which satisfy:
\begin{equation} \label{quantum_imposition_constraints}
\hat{\Omega}_A\vert \psi\rangle=0 \quad \forall A \, ,
\end{equation} 
where $\hat{\Omega}_A$ are the quantizations of all the constraints of the theory, both primary and secondary. That is, states in the physical Hilbert space are states which satisfy the quantum version of the constraint equations $\Omega _A=0$ which define the constraint surface.

Notice that the imposition of the constraints has as a consequence that states in the physical Hilbert space are invariant under the action of the quantum counter-part of the gauge generators, as they will take the form $\hat{G}=\epsilon^A\hat{\Omega} _A$ and hence $\hat{G}\vert\psi\rangle =0$. This is an important difference with the classical case: in the classical case a point on the constraint surface was not gauge-invariant, i.e., under the action of the gauge generator this point would in general change. Therefore, while the classical constrained formalism allowed for some gauge freedom, the quantum formalism only allows for gauge-invariant states. This will be a source for the problem of time for reparametrization invariant systems.

Finally, the dynamics is specified by a Hamiltonian operator defined on the physical Hilbert space. This operator can be defined by quantizing the total Hamiltonian and by noticing that the terms to $v^{\alpha}\hat{\phi}_{\alpha}$ don't have any effect on the evolution due to the conditions \ref{quantum_imposition_constraints}. In this sense, we also see how the gauge freedom and the indeterminacy that we had in the classical case has disappeared in the quantum case, as the dynamics is independent of the choice of the arbitrary functions $v^{\alpha}$ and is defined by the canonical Hamiltonian.

There are some technicalities that may arise in the quantization of a constrained system, such as the possibility of anomalies, i.e., the possibility that the relations between the classical constraints no longer hold for the quantum ones. Leaving those aside, the quantization procedure for a constrained, first-class system can be summarized as:
\begin{enumerate}
\item{Start with a classical gauge theory defined on a phase space.}
\item{Choose a subalgebra of functions on phase space and quantize them, i.e., build an algebra of operators on a kinematical Hilbert space $\mathcal{H}_{kin}$ such that their commutator algebra is defined by the Poisson algebra of the classical functions (eq. \ref{correspondence_Poisson_commutators}).}
\item{Impose the constraints. That is, define the physical Hilbert space $\mathcal{H}_{phys}$ as the space of the states which satisfy $\hat{\Omega}_A\vert\psi\rangle=0$. The states in this space are automatically gauge invariant.}
\item{Build a Hamiltonian operator which is a quantization of one of the Hamiltonians that generate the constrained dynamics in the classical theory. The dynamics of the theory is contained in equation \ref{quantum_dynamics_canonical} or in some equivalent form.}
\end{enumerate}
Next, I will show how for reparametrization invariant systems this procedure leads to the problem of time.

\subsection{Quantization of reparametrization invariant theories: the problem of time}

A theory with a reparametrization invariance is a theory which is expressed in a way that is independent of a choice of coordinatization, either of spacetime or of a trajectory or structure in a space or spacetime. The best-known example is of course general relativity, but any theory can be expressed in a reparametrization invariant way by introducing the appropriate structures, as will be clear in the next section. From a formal perspective, we can classify reparametrization invariant theories into two groups: theories with homogeneous Lagrangians and generally covariant theories. In the first category we will find theories like the two examples I will study in more detail in this paper, and they can be shown to be constrained systems which have a vanishing canonical Hamiltonian and $D$ constraints per space point in $D$ dimensions. These $D$ constraints are in correspondence with the $D$ degrees of freedom one has for choosing a set of $D$ coordinates in $D$ dimensions. The total Hamiltonian can therefore be expressed as:
\begin{equation} \label{total_ham_homogenous}
H_T=\int d^{D-1}x \left( N\mathcal{H}_0 + N^a\mathcal{H}_a \right) \, ,
\end{equation}
where there is implicit a summation over the space indices $a$, $\mathcal{H}_{0}$ and $\mathcal{H}_{a}$ are the $D$ constraints of the theory, and $N$ and $N^a$ are arbitrary functions, just as the $v^{\alpha}$ before. I will comment on the meaning of the constraints and the multipliers below, after I introduce generally covariant theories.

Generally covariant theories are theories formulated on the language of differential geometry. In this language coordinates on their own don't have any geometrical meaning, but instead the geometric properties of spacetime are encoded in the metric tensor $g_{\mu\nu}$\footnote{Here I will be using the $\{-,+,+,+,...\}$ sign convention.} which can be used to define quantities like distances and angles. Generally covariant theories are invariant under reparametrizations which take into account the geometrical nature of this formalism, that is, under reparametrizations the different geometrical objects have different transformation rules according to their geometrical properties: scalars transform as scalars, tensors as tensors, and so on.

For this type of theory we will find a similar constraint structure to the one one finds for theories with homogeneous Lagrangians. For studying a generally covariant theory in the canonical formalism it will be useful to introduce the ADM variables\footnote{The name comes from its proponents in \cite{Arnowitt:1962hi}.}. In this formulation spacetime is foliated into spacelike surfaces of constant time coordinate and the components of the metric are divided into two groups, which will play different roles. First, the spatial components of the metric $g_{ab}$ will describe a metric for the $D-1$ space which is evolving in time. The rest of the components $g_{0\mu}$ get a different interpretation: they can be seen as encoding the information about the vector $n$ normal to the spacial slice at every instant of time. A convenient way of expressing this is by means of the lapse function $N$ and shift vector $N^a$\footnote{In particular, the relation between the metric components and these functions is: $N=\sqrt{-g^{00}}$ and $N^a=-g^{0a}/g^{00}$.}. These functions allow to express the vector field $\partial _t$ which describes the foliation in terms of the normal to the foliation and a tangential vector:
\begin{equation} \label{vector_field}
\frac{\partial}{\partial t}=N\hat{n}+N^a\frac{\partial}{\partial x^a} \, .
\end{equation}
Now, when applying the canonical formalism to a generally covariant theory expressed in these variables we find a canonical Hamiltonian of the form:
\begin{equation}
H_c=\int d^{D-1}x \left( N\mathcal{H}_0 + N^a\mathcal{H}_a \right) \, .
\end{equation}
This form is suggestive, as the parallelism with \ref{vector_field} is evident: the canonical Hamiltonian which generates evolution in $t$ is naturally decomposed in two parts one which generates evolution normal to the foliation and another tangential one, with $N$ and $N^a$ describing precisely how $\partial t$ decomposes in these two components. Notice also that this is the form we found for the total Hamiltonian (\ref{total_ham_homogenous}) for a theory with homogeneous Lagrangian. In this case, however, $N$ and $N^a$ are dynamical variables, although one finds that the momenta conjugate to them, $P_0$ and $P_a$ are primary constraints of the theory. Furthermore, $\mathcal{H}_0$ and $\mathcal{H}_a$ are secondary constraints, and are known as the Hamiltonian and momentum constraints. The total Hamiltonian of a generally covariant theory can therefore be expressed as:
\begin{equation}
H_T=\int d^{D-1}x \left( N^{\mu}\mathcal{H}_{\mu}+\lambda^{\mu}P_{\mu} \right) \, ,
\end{equation}
where the $\lambda^{\mu}$ are arbitrary functions, just as the $v^{\alpha}$ before and I am introducing the more compact notation $N^{\mu}\mathcal{H}_{\mu}=N\mathcal{H}_0 + N^a\mathcal{H}_a$. This total Hamiltonian is slightly more complicated than the one for the homogeneous case, but it shares an essential feature for its later quantization, namely that the total Hamiltonian is just a sum of constraints. We will see that this is problematic for the quantization of the theory and will give rise to the problem of time, even if it is unproblematic from the classical perspective.

Before quantization, it is important to say a word about the gauge generator for a reparametrization invariant theory. The gauge generator is also a sum of constraints\footnote{For the expression for general relativity see for instance \cite{Pons2010}.} and it transforms a solution of the equations of motion to another solution with a different parametrization. For instance, in general relativity a transformation generated by this generator transforms a solution of Einstein equations to another which is diffeomorphic to it. Moreover, one can show that a particular case of gauge generator for any reparametrization invariant theory is precisely the total Hamiltonian of the theory. This just implies that a solution of the equations of motion and another one in which everything happens, say, 1 second later represent the same physical events.

However, the fact that the total Hamiltonian is a case of gauge generator has caused some confusion. For instance, Earman \cite{Earman2002-EARTMM} has argued that this fact has to make us consider that time evolution is just gauge evolution and that we should reconsider our metaphysical picture of time and of the physical content of a theory like general relativity. Some other authors \citep{Maudlin2002, PonsSalisbury, Pons2010} have rightly argued against this position. Indeed, there are two notions of gauge invariance, and once one is clear about this the conceptual trouble disappears. The way in which a reparametrization invariant theory is a gauge theory is in that two different solutions to the equations of motion can represent the same physical events but with a different labeling. This sense of gauge invariance affects a whole solution, or if you want, a whole spacetime. There is another sense of gauge invariance which is just instantaneous and it is exemplified in theories like electromagnetism. In these theories, we can consider the description of a system at a given time and gauge transformations to be transformations which just transform the description of the system at a time. In this sense, instantaneous gauge transformations are considered not to change the physical state of the system at a given time. However, a transformation generated by the total Hamiltonian of a reparametrization invariant theory is \textbf{not} a gauge transformation from this instantaneous perspective, as it transforms an instantaneous configuration of the system to another one at a later time. That is, two descriptions of the universe that differ in that in one every event happens one second earlier (with respect to some coordinate time) than in another can be considered gauge equivalent in the global sense of the term, but not in the instantaneous view of gauge, as the physical content at, say, $t=0$ will clearly be different for both descriptions.

This makes it the case that reparametrization invariance has to be treated in a different and more careful way than a gauge invariance like the one in electromagnetism. For instance, in electromagnetism the physical content of the theory at a time is in the quantities that are invariant under gauge transformations. However, imposing something like that in the case of a reparametrization invariant theory is too strong, as imposing invariance under the action of the gauge generator at a time is equivalent to imposing invariance under time evolution\footnote{This has also been argued in \cite{Pitts2017}, where it was further argued to define observables not as invariant under the action of the gauge generator but as covariant.}. This will also turn out to be problematic in the quantum case.

Finally, we can analyze what happens when we apply the steps outlined in the previous subsection to a reparametrization invariant theory. When we apply step 3 we find that states in the physical Hilbert satisfy the constraint equations for the Hamiltonian and momentum constraints:
\begin{equation}
\mathcal{H}_{\mu}\vert\psi\rangle=0 \, ,
\end{equation}
and also for the momenta conjugate to $N^{\mu}$ in the generally covariant case:
\begin{equation}
{P}_{\mu}\vert \psi\rangle=0 \, ,
\end{equation}
Now, when we try to apply the fourth step we are in trouble, as the fourth step tells us that the dynamics is defined by the total Hamiltonian of the theory, but its action on physical states vanishes. Therefore, we find that we are missing the dynamical part of our quantum theory and that we have defined a physical Hilbert space but we lack a temporal evolution equation for states in this space. This is in a nutshell the problem of time of reparametrization invariant theories.

Notice that the problem arises because the imposition of the constraints is too strong, as it plays a double role: not only it makes it the case that the constraints are satisfied but it also implies an invariance under the quantum counterpart of the gauge generator. For a gauge theory like electromagnetism this may be fine, but for a reparametrization invariant theory like general relativity it is not, as I have argued above that temporal evolution is not a gauge transformation from an instantaneous point of view.

In the next section I will show how this problem can be overcome for deparametrizable models, and how this inspired tentative solutions for the problem of time in general relativity. However, I will argue that these resolutions work precisely because of the deparametrizability of these theories. I will later argue that general relativity is not deparametrizable and that one may doubt about the applicability of these resolutions to it.

\section{Deparametrizable models} \label{deparametrizable_models}

This section is divided into two parts: first I introduce an example of a deparametrizable model and explain the way the problem of time arises in it and then I explain how there are some resolutions which are able to give us a meaningful quantum theory that overcome this problem. I argue that these resolutions rely on the deparametrazability of the model.

As it will become clear below, a deparametrizable model is a model which is defined in such a way that time, or spacetime coordinates, can be identified as variables of the model. That is, in this kind of model time (or spacetime coordinates) will appear explicitly as a variable in configuration or phase space or it can be identified and separated by means of some appropriate transformation, such as a coordinate transformation in configuration space or a canonical transformation in phase space. Evolution in these models is defined with respect to arbitrary parameters, but these models can be deparametrized, i.e., these parameters can be eliminated and the dynamics can be expressed with respect to the physically meaningful time or spacetime coordinates.

\subsection{Example: non-relativistic particle}

In this subsection I will study a simple model, the reparametrization invariant version of the dynamics of a classical particle. This simple case is commonly\footnote{For instance, this case is discussed in \cite{Isham1993,Kuchar1992,Dittrich2005,Rovelli2015}.} used as an example to follow for the quantization of general relativity and hence it is an important model to study. We start with the Newtonian action
\begin{equation}
S[x]=\int dt [\frac{1}{2}m(\frac{dx}{dt})^2-V(x)] \, ,
\end{equation}
and we introduce a reparametrization invariance in the theory by introducing an arbitrary parameter $\tau$ such that the physical time $t$ is now a configuration variable which depends on $\tau$\footnote{There is a sense in which $t$ is not treated as an ordinary configuration variable, as it is imposed that is monotonic in $\tau$. This implies that for every value of $t$ there is a unique value of $\tau$.}:
\begin{equation}
S[x,t]=\int d\tau [\frac{1}{2}m\frac{\dot{x}^2}{\dot{t}}-\dot{t}V(x)] \, .
\end{equation}
One can check that this reparametrization invariant system leads to Newton equations in the $x, t$ variables. This model has a homogenous Lagrangian and therefore one can show that it has a vanishing canonical Hamiltonian and that it satisfies the Hamiltonian constraint:
\begin{equation}\label{Hamiltonian_constraint_schrodinger}
\mathcal{H}_0=p_t+\frac{1}{2m}p_x^2+V(x)=p_t+H(x,p_x)=0 \, ,
\end{equation}
where I have introduced the Hamiltonian function $H$, which is the Hamiltonian function of the system once the reparametrization invariance is removed. This model is a one-dimensional model as it only depends on one parameter, in this case $\tau$. This model is clearly deparametrizable, as we are able to identify the variable $t$ as the time variable and given any solution of the equations of motion of the model we are able to invert the relation $t(\tau)$ and express $x(\tau)$ as $x(t)$. Let me mention that in general, we can introduce an artificial temporal or spatiotemporal reparametrization invariance to any Hamiltonian theory to obtain a reparametrization invariant theory with homogeneous Lagrangian\footnote{See for instance \cite[pp. 291-294]{Sundermeyer1980}.} by adding some extra parameters. Obviously, these models are deparametrizable, as we can always eliminate these extra parameters.

We can apply the quantization schema introduced in the last section to this system. The natural kinematical Hilbert space for this system is $\mathcal{H}_{kin}=L^2[\mathbb{R}^2,dxdt]$, that is, the space of square-integrable functions both in position $x$ and time $t$. In this representation, the constraint condition for physical states takes the form:
\begin{equation} \label{Schrodinger_constraint}
\hat{\mathcal{H}}_0\psi(x,t)=-i\hbar\frac{\partial}{\partial t}\psi(x,t)+\hat{H}\psi(x,t)=0 \, .
\end{equation}
This equation is nothing but the familiar Schr\"odinger equation of quantum mechanics. Solutions to this equation are distributional, in the sense that they are not square integrable and do not belong to the kinematical Hilbert space. This can be seen by computing the norm of such a function on this space:
\begin{equation}
\langle \psi \vert\psi \rangle=\int dt \int dx \vert\psi (x,t)\vert^2=\int dt C=\infty \, .
\end{equation}
Here I have used the fact that the spatial norm $C$ of a function satisfying Schr\"odinger equation is conserved in time. For defining the physical Hilbert space $\mathcal{H}_{phys}$ not only one needs to specify the vector space, which is the space of functions satisfying \ref{Schrodinger_constraint}, but also one needs to specify an inner product for this space. In this case we have the natural candidate:
\begin{equation}
\langle \psi_1 \vert \psi_2 \rangle= \int dx \psi_1^* (x,t_0)\psi_2(x,t_0) \, ,
\end{equation}
which is the familiar inner product used in quantum mechanics. The parameter $t_0$ is an arbitrary time parameter, given that the unitarity property of Schr\"odinger evolution makes it the case that the value of the inner product is independent of $t_0$.

As I have explained in the previous section, when we apply the last step of the quantization procedure we find a problem of time. The total Hamiltonian is just given by the Hamiltonian constraint (multiplied by a lapse function) and hence its action on physical states vanishes. In this sense, there is no evolution with respect to the parameter $\tau$. As I commented above, this is an unwanted feature, as in the classical case the temporal evolution of the system was described by the evolution of the variables $x$ and $t$ with respect to $\tau$.

Despite this, this case doesn't seem very problematic. After step 3 we obtained functions which satisfy the Schr\"odinger equation and it is straightforward to reinterpret them as the usual wavefunctions of standard quantum mechanics. Crucial for this is that we are able to identify the configuration variable $t$ as a time variable and to, consequently, treat it differently from the position variable $x$. By doing this we recover the standard quantum mechanics of a non-relativistic particle.

Notice that this solves the problem of time in that by inspecting the states in the physical Hilbert space we were able to recover a time and a unitary time evolution with respect to it. However, the time evolution we were originally looking for, that is, evolution in $\tau$ is not recovered. This is arguably not problematic in this case, as one can argue that $t$ is the physical time and that physical time evolution is evolution in $t$. But in the case of general relativity this won't be so clear, as I will argue later. Before that, I will sketch in the next section how this case and similar examples motivate several proposals for resolutions of the problem of time.

\subsection{Proposals of resolution for the problem of time} \label{proposals}

The example I just exposed motivates a few proposals of resolution of the problem of time that I will briefly introduce here. For exhaustive reviews of these in detail I refer the reader to \cite{Isham1993} and \cite{Kuchar1992}. These resolutions take as a lesson from the last example that time is in some way encoded in the configuration space of the theory and on states on the physical Hilbert space and that one just has to cleverly interpret these states to recover time and quantum mechanics.

The first proposal of resolution is just to do exactly the same as we have done for the case of the non-relativistic particle, i.e., to take states in the physical Hilbert space and interpret them as solutions of some Schr\"odinger-like equation. For making this interpretation one has to distinguish a variable as a time variable, or $D$ variables as spacetime variables for the field-theoretic case. If one is lucky, evolution with respect to this time variable will be unitary, and standard quantum mechanics can be recovered. However, this is not necessarily the case, as is illustrated by the quantization of the relativistic particle model, which leads to the Klein-Gordon equation, which is not unitary. There are further problems with this approach, like the fact that one may be able to choose different variables as time variables, leading to different and sometimes inequivalent quantum theories.

Some of the technical difficulties of this approach may be avoided if one identifies the time variable before starting the quantization process. By doing so, one can rewrite the dynamics of the theory in terms of this time variable and avoid the reparametrization invariance and the problem of time. This strategy suffers from the same multiple choice problem I just mentioned, but it doesn't suffer from the problem of unitarity.

These two approaches share an important feature: in both cases there is a deparametrization going on, i.e., in both cases one identifies the time (or spacetime) variables and is able to express the dynamics in terms of these variables. The difference is just the moment in which this identification is performed, as it can be done, technical difficulties notwithstanding, both before or after quantization. Again, I refer the reader to \cite{Isham1993, Kuchar1992} for a longer discussion of these approaches and the technical difficulties associated with them.

A similar resolution specific for the case of general relativity is the semiclassical resolution, which only aims to solve the problem of time for a subset of states in the physical Hilbert space for which a series of approximations apply. For these states, or at least for some regimes in these states, the strategy also involves a deparametrization.

A different proposal is the frozen observables resolution, which focuses on observables rather than on states. For an observable to be well-defined in the physical Hilbert space it has to commute with the constraints. This implies that its classical counterpart has to be invariant under the action of the gauge generator. I have argued above that the instantaneous view of gauge transformations is problematic for reparametrization invariant theories and too strong. Indeed, the classical quantities that satisfy this condition are constants of motion. Despite this, authors like Rovelli \cite{Rovelli1991,Rovelli1991a} argued that the physical content of a quantum theory can be recovered from these quantities which are referred to as evolving constants of motion.

Let me illustrate this with the observables for the example of the non-relativistic particle and let me set the potential $V(x)$ to be 0. In this case we can define the classical phase space function:
\begin{equation} \label{coincidence_observable}
X_T(x,t,p_x,p_t)=x+\frac{p_x}{m}(T-t) \, .
\end{equation}
It is easy to show that this function has vanishing Poisson brackets with the Hamiltonian constraint and that it is a constant of motion the value of which is the position of the particle at physical time $T$. In this sense, it is claimed that these constants of motion represent all the dynamical content of the theory. Notice that even if these quantities don't evolve with respect to the parameter $\tau$, they form a family parametrized by $T$ and if we study the evolution of $X_T$ we recover the dynamics of the particle with respect to physical time. The idea to solve the problem of time is to extend this to the quantum case, i.e., to find the operators $\hat{X}_T$ and to recover dynamics and temporal information from their evolution in $T$.

There are a few technical issues which make it difficult to find and define observables which are well-defined in the physical Hilbert space of a complicated reparametrization invariant theory like general relativity. I refer the reader to the reviews \citep{Isham1993, Kuchar1992} and also to \cite{Dittrich2005} for a discussion of those. From a conceptual point of view there are a couple of points that are worth raising.

First, one can doubt the sense in which functions like $X_T$ are genuine observables. When we start building a theory defined on a configuration space, we usually assume that the configuration variables are the quantities for which we have a physical interpretation and we can, in principle, measure or observe. All other phase-space functions can be computed from the values of these basic variables and their rates of change. For instance, we can say that the position cubed of a particle is an observable not because one can go and measure it in any straightforward way, but because we believe the position itself to be observable. Similarly, functions like $X_T$ reflect the fact that given the position of a particle at a time $t$, and knowing the equations of motion, one can compute what its position was or will be at time $T$.

Moreover, one can define quantities like:
\begin{equation}
X_{T,a,b}(x,t)=x-a+\frac{p_x}{m}(T-(t-b)) \, .
\end{equation}
These quantities also satisfy that they have vanishing Poisson brackets with the Hamiltonian constraint and hence they will also represent well-defined operators on the physical Hilbert space. However, their physical interpretation is slightly different: they do not represent what the position of the particle will be or was at a given time, but they represent the position the particle would take at time $T$ if its actual position and time at parameter time $\tau$ were shifted by $a$ and $b$. This is a kind of function which is observable in the sense that one can compute it if one observes $x$ and $t$ and knows the equations of motion, but not in any intuitive meaning.

A possible way out of this is to focus just on the observables $\hat{X}_T$ and reject any other observable defined on the physical Hilbert space. By analyzing just these observables one is lead to an interpretation very similar to the one we found by interpreting the states in the physical Hilbert space but now from a Heisenberg picture of quantum mechanics. The relation between the approaches is explored in more detail in \cite{Dittrich2005}. However, even if for the case of the non-relativistic particle things work out well, for a more general theory we can find other problems like that the evolution of the operators, if we are able to define them, may not be unitary. But more importantly for my argument in this paper is that we have had to define a set of observables which basically defined the evolution of some variables with respect to the others. This is a way of deparametrizing the theory, as at the end of the day we are finding a set of `true' observables which evolve with respect to some `true' time. In this respect, the frozen observables resolution is dependent also on the assumption of deparametrazability, and I will raise the doubt that it may not be applicable for theories like general relativity.

Finally, let me mention a different strategy championed by Rovelli which focuses not on states or operators, but on transition amplitudes\footnote{See \cite{Rovelli2004, Colosi2003}.}. Rovelli defines transition amplitudes between states in the kinematical Hilbert space by making use of the projector\footnote{This map is referred to as the projector but it is only a proper projector map for the case of Hamiltonians with discrete spectra.} $\eta$, which is a map that takes any state in the kinematical Hilbert space and returns a state in the physical Hilbert space. In this way the inner product of the physical Hilbert space defines a transition amplitude. For instance, in the case of the non-relativistic particle one defines the transition amplitude as:
\begin{equation}\label{transition_amplitude_particle}
(\eta(x_1,t_1)\vert \eta(x_2,t_2))=\langle x_1 \vert e^{\frac{-i(t_1-t_2)}{\hbar}\hat{H}}\vert x_2\rangle= K(x_1,t_1;x_2,t_2) \, .
\end{equation}
The first equality can be reached by analyzing the exact form of the projector of this theory and the expression is the expression one finds in standard quantum mechanics for the propagator $K(x_1,t_1,x_2,t_2)$. The propagator is an object which contains all the dynamical information of the theory, as it is a representation of the evolution operator. The idea of Rovelli is to generalize this to any reparametrization invariant theory and define the transition amplitude by means of the appropriate projector map $\eta$ and inner product of the physical Hilbert space.

There are some technical and conceptual difficulties in this approach, but for now let me point out that the transition amplitude thus defined can be interpreted as a propagator in a clear-cut way in the case of some deparametrizable theories like our example. That is, by identifying which variables work as coordinates for spacetime points we can interpret the transition amplitude as the propagator which defines evolution with respect to those spacetime variables. For non-deparametrizable theories, the interpretation of such an object becomes more complicated as I will argue in the next section.

Let me conclude this section by insisting on the main point for my argument: the strategies that solve the problem of time for the non-relativistic case rely on it being a deparametrizable theory, and it is therefore at least worrisome to try to apply the same strategies to non-deparametrizable theories. In the next section I will introduce an example of one such theory and argue that one cannot successfully apply the strategies defined in this section to it.

\section{Non-deparametrizable models} \label{non-deparametrizable_models}

This section follows the same structure as the previous one. In the first part I introduce an example, in this case of a non-deparametrizable reparametrization invariant model, and the way its quantization leads to a problem of time. In the second part I go through the resolutions of the problem of time outlined in the previous section and argue that they are not satisfactory for this model. This leads to the conclusion that the problem of time is not solvable for this model and that this conclusion may be extended to other non-deparametrizable models. 

\subsection{Example: two harmonic oscillators}

The example I will study in this section is the dynamics of a system of two harmonic oscillators expressed by its Jacobi action:
\begin{equation} \label{harmonic_osc_reparametrizable}
S[x,y]=2\int d\tau \sqrt{\frac{m}{2}\left(\dot{x}^2+\dot{y}^2\right)\left(E-\frac{1}{2}(k_xx^2+k_yy^2)\right)} \, .
\end{equation}
Examples similar to this were studied in \cite{Barbour1994} for supporting a Machian perspective of time\footnote{For a more recent review of Newtonian systems studied from a Machian perspective and using Jacobi's principle see \cite{Gryb2010}.}. In this view, an absolute time scale is meaningless and what is really meaningful in mechanics is the succession of configurations. This action exemplifies well this view, as the trajectories which minimize it agree in that they describe two oscillators oscillating between the same two positions, but they disagree in the values of $\tau$ they assign to each instant. This example can be generalized to consider a whole universe as described by classical mechanics. From Barbour's perspective, one can make sense of the dynamics of the universe even in the absence of the metric aspect of time, just by retaining the ordering relation it defines. In this sense a history of the universe is described by a trajectory in configuration space, and it doesn't matter the way we parametrize this trajectory.

\begin{figure}[h]%
\centering
\includegraphics[width=0.9\textwidth]{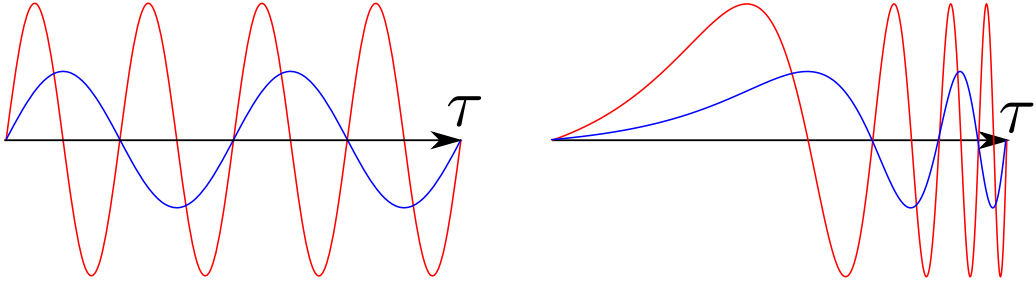}
\caption{\label{oscillators} Two equivalent solutions of the equations of motion of the double harmonic oscillator model. They represent the same sequence of oscillations but they differ from each other at the particular values of $\tau$ they assign to each moment of time. In this way the parametrization on the left-hand side represents the Newtonian parametrization in which the oscillations are regular and the one in the right-hand side represents a parametrization in which the oscillations become faster as $\tau$ passes.}
\end{figure}

Notice however that there is a choice of time parameter which is special in that it makes the equations of motion look simpler. We can define a special time parameter $t$ by imposing $dt=\sqrt{\frac{m(\dot{x}^2+\dot{y}^2)}{2E-k_xx^2-k_yy^2}}d\tau$. With respect to this parameter $t$ what one finds is that the equations of motion become the Newton equations of motion for the harmonic oscillator and that the oscillations of both oscillators are regular with respect to this time parameter, even if each one has its own frequency. In this sense, the factor $\sqrt{\frac{m(\dot{x}^2+\dot{y}^2)}{2E-k_xx^2-k_yy^2}}$ is a lapse function which plays a similar role to the one played by the metric in a generally covariant theory: it relates a physically meaningful time $t$ with the arbitrary parameter $\tau$, just as the metric relates the coordinates along a worldline with the proper time.

Systems like this one satisfy, to a degree, a thick sandwich conjecture. We can formulate the conjecture in the following way: for determining a full dynamical trajectory it is enough to specify an initial and a final state for the system, without giving any information of the time elapsed between both instants. In this case, the conjecture holds to some extent: given an initial and a final configuration one can determine trajectories in configuration space which solve the equation of motion. However, these are not necessarily unique, as, depending on the values of the initial and final states and of the couplings of the oscillators, there may be more than one trajectory in configuration space which obeys the equations of motion and satisfies the initial and final condition. Notice that this can be seen as an alternative version of classical mechanics where instead of specifying an initial and a final configuration and the time elapsed between them one specifies the configuration and the energy of the system. Therefore, there is a very natural sense in which we can say that temporal information is encoded in the energy and in the equations of motion, as the configuration space is the same in both formulations. We will find a similar conjecture for the case of general relativity in the next section.

We can now analyze this model using the Hamiltonian formulation. We find that this system is a constrained system with a vanishing canonical Hamiltonian. The total Hamiltonian is therefore:
\begin{equation}
H_T=N\mathcal{H}_0=N (\frac{1}{2m} p_x^2+\frac{1}{2m} p_y^2+\frac{k_x}{2}x^2+\frac{k_y}{2}y^2-E) \, ,
\end{equation}
where $N$ is an arbitrary positive function and $\mathcal{H}_0=\frac{1}{2m} p_x^2+\frac{1}{2m} p_y^2+\frac{k_x}{2}x^2+\frac{k_y}{2}y^2-E$ is the Hamiltonian constraint for this system. It can be shown that different choices of $N$ correspond to different parametrizations of time and that it plays the role of a lapse function. Notice that even if this example looks similar to the one in the previous section (both are theories with homogeneous Lagrangians), now time is not a variable in the configuration space of the theory. We see that this theory is not a deparametrizable one, as both degrees of freedom $x$ and $y$ represent physical degrees of freedom and not a time coordinate.

However, notice that, in a limited sense, one can use one of the variables as a clock for the other. If we consider just an oscillation of the first oscillator we can describe the position of the second one as a function of the position of the first one. That is, for a restricted amount of time we can consider $y$ as a function of $x$. But the position of the first oscillator $x$ is not a good clock for long periods of time, as there is a moment in which it reaches a maximum and turns back. Therefore specifying a value of $x$ doesn't uniquely specify an instant of time. For this reason, it is important to emphasize that even if $x$, or $y$, can work as clocks at some moment of the dynamical evolution of the system, they are not time variables. Time, in this Machian view, is defined by the sequence of configurations and not by any variable in this configuration.

Now we can consider the problem of time for this system. The kinematical Hilbert space of the system is just the space of square-integrable functions on the real plane $L^2[\mathbb{R}^2]$, and the constraint equation is:
\begin{equation} \label{constraint_eqn_double_harmonic_oscillator}
\left(\frac{1}{2m}\hat{p_x}^2+\frac{1}{2m}\hat{p_y}^2+\frac{k_x}{2}\hat{x}^2+\frac{k_y}{2}\hat{y}^2-E\right)\vert \psi\rangle=0 \rightarrow \hat{H}\vert \psi\rangle=E\vert \psi\rangle \, .
\end{equation}
This is just the time-independent Schr\"odinger equation for the system of two harmonic oscillators. Obviously, if we were expecting to get the standard quantum theory for two harmonic oscillators, the conclusion we reach is that this quantization procedure has failed. 

Notice that in the case of the non-relativistic particle we could lose evolution in $\tau$ as long as we could define evolution in $t$. In the case of the double harmonic oscillator it seems more harmful to lose evolution in $\tau$, as I have argued above that temporal evolution is a succession of configurations of both $x$ and $y$ and not definable by relative evolution as $x(t)$ was in the case of the deparametrizable model of the non-relativistic particle. In other words, as the configuration space of this example is a proper configuration space and not an extended one we don't have any time variable left if we lose $\tau$, while in the deparametrizable example we still had $t$.

Looking more carefully at what is going on in this case we find an interesting feature. In action \ref{harmonic_osc_reparametrizable} there explicitly appears $E$, which is the energy of the system. In this sense, one can read this as saying that it is part of the dynamics, or if you want, of the laws that describe the system. This is a difference with the standard formulation of mechanics, in which the total energy is a conserved quantity but it is not fixed by the action. Therefore, it is not so surprising that the quantization which stems from action \ref{harmonic_osc_reparametrizable} leads to just the quantum way of stating that the system has a fixed energy $E$. However, in quantum mechanics one needs a superposition of energies for having a non-trivial time evolution, and the Hamiltonian constraint just forbids it.

In this case, a straightforward way of resolving the problem of time is to formulate the dynamics of the two harmonic oscillators in its standard Newtonian form and quantize it. However, as we will see, for theories like general relativity this move doesn't seem to be available and hence the problem seems to be inescapable. In the next subsection I will make a few comments about how the strategies outlined in the previous section fail to solve the problem of time for our example. 


\subsection{Applying the proposals to our system}

It is straightforward to see that the strategy of deparametrizing and then quantizing doesn't really make sense if we believe our theory not to be deparametrizable. There is no variable in our phase space which represents time, and even if we were to conflate the notions of time and clock, the variables which can work as clocks only do so for a limited amount of time. More technically speaking, there is no phase-space function which increases monotonically with time, and hence no good clock for all times. Still, we can do some violence\footnote{A similar strategy can be found in \cite{Gambini2012} for a more symmetric version of the same model. The version of the model in that paper is subject to my conceptual criticisms below, although it is able to overcome some technical difficulties by adding some (somewhat ad hoc) modifications. A similar model and interpretation can also be found in \cite{Wendel2020} and it is also vulnerable to my criticisms in the main body of the text. I thank an anonymous reviewer for mentioning these papers to me.} to the formalism and replace the original constraint with:
\begin{equation}
\mathcal{H}_{0}=\frac{1}{\sqrt{2m}}p_x+\sqrt{E-\left(\frac{1}{2m}p_y^2+\frac{k_x}{2}x^2+\frac{k_y}{2}y^2\right)} \, .
\end{equation}
Phase space points satisfying this constraint automatically satisfy the original constraint, although not every point in the original constraint surface satisfies this new constraint and this constraint doesn't generate the full dynamic trajectories. This constraint has the same form as the one we found for the non-relativistic particle (\ref{Hamiltonian_constraint_schrodinger}). However, despite the formal similarity the quantization of this constraint only leads to a Schr\"odinger equation in a limited sense. The square function of an operator is only Hermitian if the operator is positive-semidefinite and it is clear that for big enough values of $x$ we get a negative number inside the square root. If the square root is not Hermitian, the evolution generated by the quantization of the constraint is not unitary. Moreover, the operator inside the square root can also become negative because of the $y$-dependent operators. Therefore, the evolution defined by the quantization of this constraint is only unitary for a limited range of values of $x$ and if we restrict the allowed states. 

Recalling the discussion before, we know that $x$ only worked as a clock for a limited range of time. In this sense one could argue that the quantum mechanics we have found is a good quantization for the second oscillator during half an oscillation of the first oscillator. However, this isn't satisfactory, as the way we are treating both variables is radically different: one is acting as a classical time parameter while the other is a quantum variable. While one oscillator just evolves classically following an oscillation, for the other we can have all the phenomena and properties typical of quantum mechanics.

The strategy of quantizing and then finding an interpretation is also problematic. For solving the quantum constraint equation \ref{constraint_eqn_double_harmonic_oscillator} it is useful to use the basis formed by harmonic oscillator energy eigenstates $\psi _n$ for both oscillators. The constraint equation in this basis becomes:
\begin{equation}
\hbar \omega_x (n_x+\frac{1}{2})+\hbar \omega_y (n_y+\frac{1}{2})=E \, ,
\end{equation}
where $\omega_x= \sqrt{\frac{k_x}{m}}$ and $\omega_y= \sqrt{\frac{k_y}{m}}$ are the frequencies for the two oscillators. This equation can be solved only for specific values of $E$. For these values, we can find values of $n_x$ and $n_y$ which satisfy the constraint. States in the physical Hilbert space are superpositions of states with those values of $n_x$ and $n_y$:
\begin{equation}
\psi (x,y)=\sum _{\hbar \omega_x (n_x+\frac{1}{2})+\hbar \omega_y (n_y+\frac{1}{2})=E}c_{n_xn_y}\psi_{n_x}(x)\psi _{n_y}(y) \, ,
\end{equation}
where $c_{n_xn_y}$ are the amplitudes for the allowed combinations of $n_x$ and $n_y$. Now, for applying the deparametrization strategy we should pick one of the variables as representing time and the other one as the dynamical variable. Obviously, the symmetry between $x$ and $y$ makes it the case that none of the choices is natural. Furthermore, if we consider $x$ (similarly for $y$) to be a time parameter we find again the problem that evolution in $x$ isn't unitary and that the amplitude for very big $x$ goes to 0. In this sense, deparametrizing after quantization doesn't do better than deparametrizing before quantization.

Now we can analyze the frozen observable strategy by studying the operators defined in the physical Hilbert space of the system. In this case, it is easy to find that the operators acting on the $x$ space commute with the constraint only if they commute with the $x$-Hamiltonian, $\frac{1}{2m}\hat{p}^2_x+\frac{k_x}{2}\hat{x}^2$, and similarly for operators acting on the $y$ space. Therefore, any function $f(n_x,n_y)$ defines an operator on the physical Hilbert space, which acts as:
\begin{equation}
\hat{f}(n_x,n_y)\psi (x,y)=\sum _{\hbar \omega_x (n_x+\frac{1}{2})+\hbar \omega_y (n_y+\frac{1}{2})=E}c_{n_xn_y}f(n_x,n_y)\psi_{n_x}(x)\psi _{n_y}(y) \, .
\end{equation}
It is unclear in which way, if any, one could find a sense of evolution which made physical sense in the family of functions $f(n_x,n_y)$.

As I said above, by looking at just half an oscillation of the first oscillator $x$, it is a well-defined question to ask what is the value of $y$ when $x$ takes a given value $X$. We can even construct a phase space observable $Y_X$ which would be analogous to $X_T$ in the example of the previous section. However, this observable would only be well-behaved for the oscillation in which we are defining it. Outside of that domain the function will not have a vanishing Poisson bracket with the constraint and won't be a constant of motion any more\footnote{The reason for this is that the inverse trigonometric functions needed to define such an observable are multivalued.}.Therefore, we have trouble with $Y_X$ even at a classical level. At a quantum level, things are technically even more complicated as the function $Y_X$ is not trivial and nothing grants us that its quantization will commute with the quantum constraint. That is, it may be the case that there is no quantization of $Y_X$ which is a well-defined operator in the physical Hilbert space.

Apart from these issues, the general worry in this section remains also for this strategy. Choosing a preferred variable to be the one which defines temporal evolution seems wrong in this case, not only because both variables are on a par, but also because we have the intuition from the classical theory that temporal evolution is defined as a sequence of configurations and not by any physical variable acting as a clock. 

Despite this, let me mention that authors like Rovelli \cite{Rovelli1991,Rovelli1991a} don't want to commit to any choice of preferred variable in their resolution of the problem of time. Therefore they would allow both $X_Y$ and $Y_X$ to contain the physical information of the theory. However, this leaves us with no clear notion of temporal evolution, as both evolution in $X$ and evolution in $Y$ would be described by some set of operators. How to make those temporal evolutions compatible is unclear. Therefore, even if we were able to define the observables, we would be left with a series of observables with an unclear interpretation which goes against our intuitions of what we expected of a quantum theory. To be fair to Rovelli, he acknowledges\footnote{Again, I refer the reader to \cite{Rovelli1991, Rovelli1991a}.} that in his view the usual Schr\"odinger picture can only be recovered for some systems or as an approximation. However, in my opinion this doesn't give a satisfactory way of interpreting the formalism and it doesn't answer the general worries about the relational strategy for non-deparametrizable models I have raised in this section.

Finally, the transition amplitude strategy is also not satisfactory for this case. One can compute the transition amplitude for two (improper) states in the kinematical state to find:
\begin{equation}
(\eta(x_1,y_1)\vert \eta(x_2,y_2))=\sum _{\hbar \omega_x (n_x+\frac{1}{2})+\hbar \omega_y (n_y+\frac{1}{2})=E}\psi ^*_{n_x}(x_1)\psi^* _{n_y}(y_1)\psi_{n_x}(x_2)\psi _{n_y}(y_2) \, .
\end{equation}
This object cannot be interpreted as a time evolution operator in any straightforward way. For instance, if we try to interpret it as an evolution operator in $x$ it is not unitary, as we find that the norm $\int dy_1 \vert (\eta(x_1,y_1) \vert \eta(x_2,y_2))\vert^2$ depends on $x_1$. That is, if we were to use $(\eta(x_1,y_1)\vert \eta(x_2,y_2))$ to define a state in $y$ we would find that the norm of this state would depend on the `time' $x$. 

As was the case for the other resolutions, the transition amplitude strategy finds the problems that we lack a time parameter for interpreting the objects we have found as describing something evolving and that taking one of the physical degrees of freedom to be something like a time not only is technically problematic, but it goes against the spirit of the classical theory.

One argument in favor of the transition amplitude strategy for the case of general relativity relies on the sandwich conjecture for that theory. The argument could also be applied for our model: in the same way that for our classical model we can know the time elapsed between an initial and a final configuration, maybe specifying an initial and a final state for the transition amplitude also determine the time for the quantum case. However, in the transition amplitude we have found there was no hint of that and the thought may be misled, as in the classical case it wasn't the initial or final configurations that carried information about time, we were able to get time back just because we knew the equations of motion and the energy of the system.

Similarly to the case of the frozen observables strategy, one may insist on the transition amplitude strategy for theories like general relativity, even if acknowledging some of the difficulties pointed out in this section. This position has been also defended by Rovelli\footnote{See \cite{Rovelli2004, Colosi2003}}. However, the physical interpretation of these transition amplitudes is at best dubious. The reason for this is that the properties of a quantum propagator do not apply to these amplitudes, and hence one cannot use these transition amplitudes for obtaining a wavefunction evolving in time, which makes it the case that one cannot apply neither the standard formalism of quantum mechanics nor any of the standard realist interpretations of the theory.


Before closing this section, let me comment that a very similar example has been discussed in the literature. This is the model introduced in \cite{Rovelli1990} and which has been discussed since \cite{Gambini2001,Colosi2003}. This model is a more symmetric version of the double oscillator model I have been considering, and a similar analysis applies to it. However, let me mention that it is a trickier case, as the symmetry makes it the case that there is a conserved quantity (the angular momentum if one reinterprets the positions of the two oscillators as the $x$ and $y$ coordinates of a single particle) and one can define a monotonically increasing phase space variable (the polar angle $\phi$). One can avoid a number of the problems raised above by considering that $\phi$ acts as a time variable and by deparametrizing the theory. This solves the technical problems, but not the conceptual ones (such as why would we want to consider $\phi$ as a time coordinate and not want to quantize it). In any case, the example I have considered in this section is generally non-deparametrizable, and hence quantizations of the symmetric model which interpret it as deparametrizable\footnote{In particular, I am referring to the quantum model developed in section VI of \cite{Gambini2001}. I thank an anonymous reviewer for pointing to me this possible counterexample.} do not invalidate my analysis of the system analyzed in this section.

Let me close this section by insisting that the problem of time for non-deparametrizable models like the one I have just analyzed seems a serious problem and that the resolutions which are popular in the quantum gravity literature don't seem to be able to satisfactorily solve it. Next, I will argue that general relativity is not deparametrizable and that the same analysis applies to it.

\section{General relativity is not deparametrizable} \label{gr_not_deparametrizable}

If the analysis of the previous two sections is correct, the deparametrazability or not of a reparametrizable theory is crucial for solving the problem of time that arises when quantizing it. We have seen several proposals of resolution which work quite well for simple deparametrizable models, not so well for more complex ones, and probably not at all for non-deparametrizable ones. Therefore, as we are interested in quantum gravity, i.e., quantizations of general relativity, we should consider whether it is a theory that can be deparametrized or not. In this section I will introduce general relativity and I will argue that it is not deparametrizable and that there is no clear sense in which its configuration variables carry information about time. 

General relativity is a case of a generally covariant theory as I have introduced them in section \ref{Problem_of_time}. As I explained in that section, to canonically quantize the theory we need to express it in the Hamiltonian formalism. However, notice that not every model of general relativity can be expressed in this way, as general relativity allows for non-globally hyperbolic spacetimes, i.e., for spacetimes which cannot be decomposed as a space evolving in time. These spacetimes are usually regarded as problematic or unphysical, as they can show features like having closed timelike curves. For this reason, the restriction to globally hyperbolic spacetimes needed for expressing general relativity in the Hamiltonian formalism isn't considered troublesome. 

Expressed in the ADM variables, a model of general relativity describes the evolution in time of a $D-1$-metric $g_{ab}$, a lapse function $N$ and shift vector $N^a$ and, possibly, some matter fields $\phi_{\alpha}$. All these quantities are defined as fields on a $D-1$-dimensional space manifold. General relativity has a reparametrization or diffeomorphism invariance, as different models can describe the same physical spacetime but described using a different foliation or a different way of assigning coordinates to spacetime points. In the discussion below I will focus on general relativity for four spacetime dimensions, that is, $D=4$.

The time coordinate describing the foliation doesn't have any metrical meaning, it just contains ordering information, i.e., leaves of the foliation with a bigger parameter go after the leaves with a smaller one. In this, it is perfectly analogous to the parameter $\tau$ in the two examples I have studied in the previous sections. However, when we want to recover metrical time information, that is, information not about which event goes before another, but information about how much time has elapsed between the two events, what we find is that general relativity is similar to the example in the last section and not to a deparametrizable theory. In the deparametrizable theory, metrical time was one of the variables in the configuration space of the theory, while in general relativity none of our variables seems to be metric time. In the case of the two harmonic oscillators time wasn't a variable in the configuration space and the dynamics was independent of the way we chose to label time, but there was a preferred time we could define which was no other than the Newtonian time. This time was defined by:
\begin{equation}
dt=\sqrt{\frac{m(\dot{x}^2+\dot{y}^2)}{2E-k_xx^2-k_yy^2}}d\tau \, .
\end{equation}
This definition can of course be generalized to any Newtonian system, as shown by Barbour\footnote{I refer the reader again to \cite{Barbour1994} for a discussion of Newtonian systems using Jacobi's principle.}. In general relativity we seem to be in the same situation as the proper time between two events is defined infinitesimally as:
\begin{equation}
ds^2=-g_{\mu\nu}dx^{\mu}dx^{\nu}=N^2dt^2-g_{ab}(dx^a+N^adt)(dx^b+N^bdt) \, .
\end{equation}
Proper time in general relativity and Newtonian time in the example in the previous section are analogous according to Barbour. In both cases we have well-defined dynamical theories which determine a set of physical events with precise ordering relations\footnote{These ordering relations are different in both cases. In the Newtonian case the ordering relation is given by the absolute time, while in the general relativistic case events are partially ordered by the causal structure of spacetime.} and, even if we can define a preferred metric time in both cases, it is not necessary to do so. In this sense, Barbour considers Newtonian time in Newtonian physics and proper time in general relativity to be convenient ways of treating time that may simplify our calculations, but he rejects that they have any further meaning as the `true' scale of time. 

In both cases we can define an ideal clock as a system which directly correlates its physical state with the metric time: in the Newtonian case the reading of an ideal clock gives the absolute time, while in the general relativistic case an ideal clock shows the proper time elapsed along its worldline. Of course, real clocks are not ideal and there may not exist any real physical system which ever measures metric time. In any case, it should be clear that in both models metric time and clocks are defined in analogous ways. Therefore, there is a good case supporting that time in general relativity is represented in a similar way to the one chosen by Barbour and not to the one in the deparametrizable models. Notice that this conclusion holds independently of our philosophical position regarding absolute time scales.

Let me now introduce the phase space structure of general relativity\footnote{I refer the reader to \cite{DeWitt1967b,Arnowitt:1962hi,Pons2010,Isham1993} for more detailed introductions and derivations of the canonical formulation of general relativity. These derivations and formulations are based on the Einstein-Hilbert action of general relativity and are of extended use in the literature. Let me mention however that general relativity can also be expressed by means of the BSW action and that one can perform a canonical analysis based on this formulation. Expressing general relativity in that way wouldn't affect my analysis, and it has been used to make emphasis on the connection with Jacobi-like actions which is a point I am also arguing for in this paper. I refer the interested reader to \cite{Barbour2002} for a discussion of general relativity in terms of the BSW action.}. First, the momenta conjugate to the spatial metric are:
\begin{equation}
\pi^{ab}=-\vert g\vert ^{1/2}\left( K^{ab}-g^{ab}K\right) \, ,
\end{equation} 
where $K^{ab}$ is the extrinsic curvature tensor which describes how the spatial metric changes along the direction normal to the space surface, that is:
\begin{equation}
K_{ab}= \mathcal{L}_{n}g_{ab}\, .
\end{equation}
The momenta conjugate to the lapse function and shift vector are primary constraints with no physical meaning, as discussed above:
\begin{equation}
P_{\mu}=0 \, .
\end{equation}
When matter fields are present, we will also have some momenta $\pi^{\alpha}$ associated to the fields. As a generally covariant theory it has the same constraint structure I have described above, with Hamiltonian and momentum constraints given by:
\begin{eqnarray}
\mathcal{H}_0= G_{abcd}\pi^{ab}\pi^{cd}-\vert g\vert ^{1/2}R[g]    + \mathcal{C}_{0, matter}      \label{Hamiltonian_constraint_gr}   \\
\mathcal{H}_a =-2\nabla _b  \pi _a^b + \mathcal{C}_{a, matter} \, , \label{momentum_constraint_gr}
\end{eqnarray}
where $\mathcal{C}_{\mu,matter}$ are the matter contributions to the constraints, $R$ is the 3-dimensional curvature of space, and $G_{abcd}$ is known as the supermetric and defined as:
\begin{equation} \label{supermetric}
G_{abcd}=\frac{\vert g\vert^{-1/2}}{2}\left( g_{ac}g_{bd}+g_{bc}g_{ad}-g_{ab}g_{cd}     \right) \, .
\end{equation} 
These constraint equations correspond to 4 of the Einstein equations of general relativity, namely to:
\begin{equation}
G^{0\mu}=\kappa T^{0\mu} \, ,
\end{equation}
where $G^{0\mu}$ are the temporal components of the Einstein tensor, $T^{0\mu}$ are the temporal components of the stress-energy tensor in the case we consider matter fields and $\kappa$ is a constant. These 4 equations are constraint equations, that is, they don't determine how the variables of the theory evolve, but they just are conditions that have to be satisfied at every instant of time. This means that in general relativity one cannot choose arbitrary initial conditions, just as in electromagnetism one has to require of the initial state of the electric field that it satisfies Gauss law.

The other 6 Einstein equations are dynamical equations and they can be derived in the constrained Hamiltonian formalism from Hamilton equations for the total Hamiltonian:
\begin{equation}
H_T=\int d^{3}x \left( N^{\mu}\mathcal{H}_{\mu}+\lambda^{\mu}P_{\mu} \right) \, ,
\end{equation}
where the 4 arbitrary $\lambda^{\mu}$ reflect the formal indeterminism of the theory, as different choices of them would correspond to different coordinatizations of the spacetime compatible with the initial conditions.

We see that the canonical analysis of general relativity is what we were expecting from our general analysis in section \ref{Problem_of_time} and from the other examples studied in this paper. Therefore, its quantization will lead to a problem of time, and hence it will be important to determine whether it is a deparametrizable theory or not. The main difference with the other examples, at the level of the canonical formalism, is that in general relativity we have secondary constraints, and the variables $N$ and $N^a$ are a bit special. Their equations of motion are determined by the arbitrary $\lambda^{\mu}$ and therefore we have some freedom for choosing them in any way that is convenient. In this sense they are more like Lagrange multipliers and when considering the initial value formulation of general relativity one can focus on the spatial metric and matter fields.

The remaining elements of the phase space are the metric $g_{ab}$ and matter fields $\phi_{\alpha}$, and their conjugate momenta $\pi^{ab}$ and $\pi^{\alpha}$. As I commented before, general relativity doesn't look like a deparametrizable theory, as all of these phase space variables have their own physical interpretation and don't seem to represent a time variable. In particular, the interpretation of matter fields is clear, and given that we can consider general relativity for different kinds of matter fields and even for no matter fields at all, if general relativity were a deparametrizable theory it seems that the most natural place to look for spacetime variables is in the geometric degrees of freedom, that is, in $g_{ab}$ and $\pi^{ab}$\footnote{I thank an anonymous referee for raising the objection that general relativity may become deparametrizable by adding the matter degrees of freedom, as suggested for instance by the model in \cite{Brown1995}. However, there are several problems with this line of objection. First, these additional matter fields that need to be added are pretty special and there is nothing in our standard model of particle physics similar to them. Furthermore, these fields may be unphysical, as they may need to violate the energy conditions of general relativity (see for instance the discussion in \cite{Giesel2015}). From a more conceptual point of view, even if the fields behaved in a monotonic way in time and space, one could resist the claim that the theory is deparametrizable, as this claim seems to conflate the notions of clock and time, and the notions of rods and space. Therefore, and for the reasons above, I will not consider the matter content of general relativistic models in order to assess their deparametrizability.}. A way of formulating this is the following. We would be looking for a canonical transformation of the form:
\begin{equation} \label{canonical_deparametrization}
g_{ab}, \pi^{ab} \rightarrow X^{\mu}, p_{\mu}, \phi^{A}, \pi _A \, .
\end{equation} 
That is, we are looking for a transformation that separates the 12 phase space variables (6 components of the metric and 6 conjugate momenta) into two groups. In the first group we would have 4 coordinates $X^{\mu}$ that would be able to identify any spacetime point and 4 momenta $p_{\mu}$ conjugate to them. And the second group would represent the true dynamical degrees of freedom of general relativity, which would be contained in two fields $\phi^{A}$ and conjugate momenta $\pi _A$. This would be very attractive, as it would allow us to express general relativity in terms of the physical coordinates $X^{\mu}$, it would give us a clear picture of the physical content of general relativity and would allow us to apply the quantization strategies outlined in section \ref{deparametrizable_models}. Moreover, according to this canonical transformation there would be only two true gravitational degrees of freedom $\phi ^A$, in accordance with other arguments from general relativity that have led physicists to believe that such is the number of degrees of freedom of general relativity. For instance, it is a well known fact that there are two possible polarization states for gravitational waves.

However attractive this proposal may sound there are two problems with it. First, at a conceptual level we have seen that the variables of general relativity have a physical meaning on their own and none seems to be encoding time. Therefore, it doesn't look possible that one could build spacetime coordinates from them. Consider again the analogy with the model of the two harmonic oscillators: it seems quite obvious that in the configuration space of the two harmonic oscillators we have just the description of the two oscillators and nothing else. Similarly, in the configuration space of general relativity what we have is the description of a three-geometry (and maybe some matter) and nothing else. Second, there is a major technical difficulty: Torre \cite{Torre1992} showed that the constrained phase space of general relativity cannot be identified with the constrained phase space structure of a deparametrizable model\footnote{For a deparametrizable model, if we perform a transformation like \ref{canonical_deparametrization}, we find that the constraints take the form $p_{\mu}+h_{\mu}(X^{\mu}, \phi ^A, \pi _A)$ for some functions $h_{\mu}$. This can be shown to imply that the constrained space is a manifold, i.e., that it satisfies some properties like being smooth. Torre showed that the constrained space in the case of general relativity is not smooth, and hence that the constraint spaces cannot be identified.}, and hence it cannot be treated as such. 

This analysis already shows that general relativity is not a deparametrizable theory. Nevertheless, in the quantum gravity literature\footnote{See for instance the quotations below from \cite{Kiefer2012}.} one can find claims that time is hidden in the configuration space of the theory. In this way, the hope is that general relativity could be something in between both examples in this paper: a non-deparametrizable theory but which still has information about time in its configuration space and for which the resolutions for the deparametrizable models could apply. However, the way time is supposed to be hidden in general relativity is unclear and I will now argue against the three arguments most commonly used for supporting that claim.

The first argument comes from the thick sandwich conjecture of general relativity, first stated in the 1962 paper ``Three-dimensional Geometry as Carrier of Information about Time'' \citep{Baierlein1962}. In this paper it is argued that given the thick sandwich conjecture, one can see three-geometry as carrying some information about time. The sandwich conjecture in this case is similar to the one I have introduced for the system of two oscillators: given an initial and a final 3-geometry the equations of motion of general relativity uniquely determine the full 4-dimensional geometry in between, without needing to specify any information about the time elapsed between the initial and final states. However, even if the conjecture turned out to be true, this doesn't mean that the 3-geometry carries information about time, as it should become clear if we compare with the example of the two-harmonic oscillators. It is only by means of the equations of motion of the system that the initial and final configurations of the oscillators determine the time elapsed between the initial and final moments and hence I rejected that the configuration of the oscillators carried information about time. Similarly, it is not the 3-geometries, but the Einstein equations which would determine the temporal information in the case of general relativity, were the conjecture to be true.

The other argument supporting that time is somehow included in the configuration space of general relativity comes from some counting of degrees of freedom. This kind of argument performs an analysis of either the configuration or phase space of general relativity, finds that there are more degrees of freedom than the two physical degrees of freedom that there are believed to be in general relativity and argues that the difference has to be in the temporal information that 3-geometries are supposed to carry. For instance, Kiefer \cite{Kiefer2012} provides two ways of reaching this conclusion. First, in configuration space
\begin{quote}
The three-metric $h_{ab}$[$g_{ab}$ in my notation] is characterized by six numbers per space point (often symbolically denoted as $6\times\infty^3$). The diffeomorphism constraints (4.70) [Momentum constraints \ref{momentum_constraint_gr}] generate coordinate transformations on three-space. These are characterized by three numbers, so $6-3 = 3$ numbers per point remain. The constraint (4.69) [Hamiltonian constraint \ref{Hamiltonian_constraint_gr}] corresponds to one variable per space point describing the location of $\Sigma$ in space-time (since $\Sigma$ changes under normal deformations). In a sense, this one variable therefore corresponds to ‘time’, and $2\times\infty ^3$ degrees of freedom remain.
\cite[p. 114]{Kiefer2012}
\end{quote} 
Second, in phase space:
\begin{quote}
[...]the canonical variables $(h_{ab}(x),p^{cd}(y))$ are $12\times\infty^3$ variables. Due to the presence of the four constraints in phase space, $4\times\infty^3$ variables have to be subtracted. The remaining $8\times\infty^3$ variables define the constraint hypersurface $\Gamma _c$. Since the constraints generate a four-parameter set of gauge transformations on $\Gamma _c$ (see Section 3.1.2), $4\times\infty^3$ degrees of freedom must be subtracted in order to ‘fix the gauge’. The remaining $4\times\infty^3$ variables define the reduced phase space $\Gamma _r$ and correspond to $2\times\infty^3$ degrees of freedom in configuration space —in accordance with the counting above.
\cite[pp. 114-115]{Kiefer2012}
\end{quote}  
This kind of reasoning works well for gauge theories like electromagnetism. However, there are reasons to doubt that it may also apply to general relativity. In the first place, I have argued above that reparametrization invariance cannot be treated exactly as a gauge theory, as a reparametrization is a gauge transformation from the global point of view, i.e., it transforms solutions of the equations of motion to physically equivalent solutions of the equations of motion, but not from the instantaneous point of view: given two reparametrization-equivalent models, the physical state at the instant represented by a given parameter time is in general different for both cases. Second, to speak about the degrees of freedom at a spacetime point is tricky in the case of a generally covariant theory: while in the case of a theory like electromagnetism for specifying a spacetime point it is enough with giving the coordinate point, in the case of general relativity to speak about the degrees of freedom at spacetime point is harder because the transformations we care about move things around, i.e., change the coordinate points where physical events happen and also because the concept of spacetime point is harder to define.

To see that the degree of freedom counting is unreliable, we can try to apply it to the examples in the previous sections.  Both cases are formally analogous: we have configuration spaces with two variables $x,t$ or $x,y$, both are described by homogeneous Lagrangians and in both cases evolution and gauge are generated by the Hamiltonian constraint. We can count the degrees of freedom applying Kiefer's method. In configuration space we get that there supposedly is 1 degree of freedom in both theories, that is 2 degrees of freedom - 1 gauge transformation. Similarly, in phase space we start with 4 variables and we have to subtract 1 constraint and 1 gauge transformation, giving the two phase space variables associated with one degree of freedom. For the case of the non-relativistic particle we are satisfied with the outcome as the system we are describing has one degree of freedom, namely the position of the particle.

However, for the case of the double harmonic oscillator the result of the counting seems wrong. The system we are describing is a system formed by two oscillators, and intuitively it has two physical degrees of freedom. The conclusion of the counting of degrees of freedom argument for this case would lead us to think that of the two oscillators one would be something like a time and the other one the physical degree of freedom. Above I have argued that even if we can use one of the oscillators as a clock, that is, as a device to keep track of time, it doesn't make sense to consider it as a time and the other as something like the real physical system. One could be more sophisticated and argue that time is not directly one of the configuration variables but some combination of them or that the way time is encoded in the configuration variables is not so straightforward. However, as I have been arguing this doesn't seem very plausible and it seems more reasonable to state that the configuration variables are just configuration variables and that the only way in which they carry information about time is by means of the equations of motion. Therefore, we should reject the degree of freedom counting argument for this case.

This shows that one should be careful when applying this sort of reasoning. Moreover, given the strong analogy between the double harmonic oscillator example and general relativity I have been arguing for, we have a good case for rejecting the conclusion from Kiefer's arguments that in general relativity time is encoded in three-geometry. This rejection raises doubts about the applicability of the resolutions of the problem of time to the case of general relativity. 

Finally, the third kind of argument which is used in the literature for claiming that time is part of the configuration or phase space of general relativity is by making more or less explicit proposals for this identification\footnote{I am thankful to Brian Pitts for pointing out this kind of argument to me.}. In \cite{Kuchar1993} some of this proposals are explored and found to have some conceptual problems. In particular, let me mention one particularly strong proposal, which is suggested by the form of the Hamiltonian constraint \ref{Hamiltonian_constraint_gr}. The first term in this expression, $G_{abcd}\pi^{ab}\pi^{cd}$ is known as the kinetic term and it is formally analogous to the kinetic term one finds in the quantization of a relativistic particle, $\eta_{\mu \nu}p^{\mu}p^{\nu}$. In this case, the momenta $p^0$ are conjugate to the time variable, and one can characterize them because they are time-like with respect to the Minkowski metric $\eta$, i.e., they satisfy $\eta_{\mu\nu}p^{\mu}p^{\nu}<0$ . In the case of geometrodynamics we find that the supermetric $G_{abcd}$, as defined in \ref{supermetric}, is hyperbolic with signature $\{-,+,+,+,+,+\}$. That is, the supermetric is analogous to the Minkowski metric in that it defines a `time-like' direction in superspace, the space of 3-metrics $g_{ab}$. This suggests that time is encoded in this `time-like' direction, just as in the case of the relativitic particle time is `marked' in the constraints by a negative sign.

However, there are several reasons to resist this kind of argument. First, the supermetric is a metric in an abstract infinite-dimensional space and its relation, if there is to be one, with the metric of spacetime is unclear. Kuchar \cite{Kuchar1993} points this out, and he argues that even for Euclidean spacetimes, i.e., even for spacetimes with no time-like direction, the supermetric would still have the same hyperbolic signature. In this sense, the fact that the supermetric has a `time-like' direction seems to be just a consequence of the form of the dynamics and not related with the structure of spacetime. Second, the time-like direction of the supermetric is associated with conformal transformations, i.e., with transformations which just (locally) expand or contract space. However, Kuchar also argues that identifying something like a local volume element with the time variable is problematic, as we can conceive of spacetimes that in their evolution expand, contract or remain at a fixed volume. In this sense, from a conceptual point of view the identification of some component or function of the metric tensor with time remains problematic.

Let me also complement those arguments with a comparison with the double harmonic oscillator model. In this model we don't have a kinetic term with negative sign but we can introduce a change to the model to introduce a negative sign. The action would now become:
\begin{equation} \label{harmonic_osc_reparametrizable_negative_sign}
S[x,y]=2\int d\tau \sqrt{\frac{m}{2}\left(-\dot{x}^2+\dot{y}^2\right)\left(E-\frac{1}{2}(k_xx^2+k_yy^2)\right)} \, .
\end{equation}
The equations of motion for such an action do not represent Newtonian trajectories, and they could be considered unphysical. Despite this, the interpretation of the model remains the same: it represents the evolution of two degrees of freedom with respect to a parameter $\tau$. Now we have an asymmetry between $x$ and $y$, and we could take the minus sign in the kinetic term for $\dot{x}$ to signal that $x$ has become time. However, this would be wrong, as just changing the form of the dynamics doesn't change the interpretation we make of the variables and what they represent. Moreover, the dynamics of $x$ still allow for $x$ to behave in a not monotonically way: even if solutions to the equations of motion do not correspond with an harmonic oscillator any more, some solutions still describe trajectories with velocities which change sign. This example shows that a negative sign in the action\footnote{This negative sign is also translated to a negative sign in the Hamiltonian constraint: $H_0=-\frac{1}{2m} p_x^2+\frac{1}{2m} p_y^2+\frac{k_x}{2}x^2+\frac{k_y}{2}y^2-E$} is not necessarily a hint that some variable represents time, as this negative sign can arise naturally in some non-Newtonian models, and it doesn't imply that that variable will behave monotonically. In this sense, in general relativity it seems plausible that the negative signs that arise are just a consequence of the form of the dynamics and it may be wrong to import our intuitions and interpretations from other models. Furthermore, as I have argued above, to argue that something like a local volume element is time one would need to show that it behaves monotonically. And even if that case, one could still argue against this identification, as from an intuitive point of view this volume element could be argued to be a configuration variable with a physical meaning and not time. 

With this I conclude this section, where I have argued that general relativity is not deparametrizable and that the arguments supporting that time is encoded in the configuration or phase space of the theory are misleading, as, among other reasons, they would lead to wrong conclusions for the case of the double harmonic oscillator. In this sense, the arguments in this section show the strength of the analogy between general relativity and the double harmonic oscillator example. In the next section I will comment on the consequences this has for the quantization of general relativity.

\section{Consequences for quantum gravity} \label{consequences}

My argument above should have made it clear that the non-deparametrazability of general relativity means that the problem of time for any canonical quantization is more worrisome than for a model like our first example and that there are good grounds for believing that the resolutions that worked for that example won't work for this case. In this section I will make a few comments on the consequences this has for different approaches to quantum gravity.

First, we have quantum geometrodynamics which is the direct quantization of general relativity in its ADM formulation as I have formulated above. That is, in geometrodynamics one seeks for wavefunctions defined on superspace, that is the space of 3-dimensional metrics $g_{ab}$. It should be straightforward to see that my arguments above directly affect this attempt to formulate a quantum theory of gravity. For instance, Kiefer \cite{Kiefer2012} argues for a semiclassical resolution of the problem of time in the context of quantum geometrodynamics, but this resolution relies on the assumption that one can interpret some of the variables in the configuration space as a time variable, but the argument above give us strong reasons for believing that that is not the case. For further criticisms of semiclassical approaches see \cite{Chua2021}.

After the discovery of the connection formulation of general relativity by Ashtekar\footnote{See \cite{Ashtekar1986}.} part of the quantum gravity community shifted to loop quantum gravity (LQG)\footnote{See \cite{Rovelli2004} for the standard reference in LQG.}, which is the result of applying the canonical quantization program to general relativity expressed in this new set of variables. In this formulation the information that before was encoded in the 3-metric $g_{ab}$ is represented by the triad field $e^a_i$ which can be thought of as a set of orthonormal vectors at every point of the three-manifold. The conjugate variable to the triad field is the connection $A^i_a$, which is also related to the extrinsic curvature $K_{ab}$. The connection variables contain essentially the same information as $g_{ab}$ and $K_{ab}$, together with some extra gauge freedom, but their introduction allows for some simplifications that motivated the shift from the purely geometrodynamical variables to them. However, from the perspective of the problem of time nothing really changes with the introduction of these variables: we still have a reparametrization invariant theory that shows a problem of time when quantizing. Furthermore, the same arguments that lead us to think that general relativity is not deparametrizable and that the strategies for solving the problem of time in deparametrizable theories won't work for general relativity will also apply for this reformulation of the theory. 

Let me mention the two ways in which the problem of time has most commonly been addressed by the LQG community. The first way was by means of the Dirac observables strategy that I have outlined before. For instance, in the book \cite{Thiemann2007} this strategy is advocated. The way this is done is exactly as described in \ref{proposals}, i.e., the observables Thiemann aims to define are coincidence observables which describe the value that a physical quantity takes when others (4 in the case of general relativity) take some values, just like $X_T$ in \ref{coincidence_observable} described the position of the particle at the time $T$. However, as I have argued above, this strategy doesn't seem to work, not only for technical but also and more importantly for conceptual reasons, if the theory we are considering is not deparametrizable, just as in the case of general relativity. In this sense, my arguments above lead one to reject a formulation of LQG based on the Dirac observables strategy like Thiemann's.

More recently, the LQG community has shifted to the transition amplitudes strategy and to formulations of the theory which rely on covariant quantizations like the ones used in spin foam models. This shift is well-represented by Rovelli's book \citep{Rovelli2004}, where it is explicit that the goal of the theory is the definition of `transition amplitudes' in an analogous way to the transition amplitudes I defined before in equation \ref{transition_amplitude_particle}, that is, the transition amplitude is defined using states in the kinematical Hilbert space, a map $\eta$ to the physical Hilbert space and an inner product in this space. In covariant formulations and spin foam models one can define transition amplitudes by means of some analog of a path integral without referring to the canonical formalism, but notice that it is usually the case that some equivalence between the formalisms is expected\footnote{Consider for instance the textbook \cite{Rovelli2015}. The transition amplitudes defined in this book (see for instance chapter 7) are defined by means of a spin foam model and not directly from the canonical formalism, but they are nevertheless considered to be some sort of approximation to the `true' transition amplitudes which would agree with the ones defined by the canonical formalism (see the discussion in sections 2.4 and 8.3.2).}. Even if that were not the case, some of the conceptual worries that I have raised here about the transition amplitude still apply to transition amplitudes defined in ways different from the canonical formalism.

To insist, the transition amplitudes strategy worked well for the case of the deparametrizable model of the non-relativistic particle as the transition amplitudes it defined are nothing but the propagator of the theory, which allowed one to recover the standard formulation of quantum mechanics. For non-deparametrizable models I have argued that the quantity defined as a transition amplitude does not satisfy the properties we would expect of a propagator and hence it cannot be given that interpretation\footnote{Interestingly, for the case of LQG the same point is raised in \cite[p. 96]{Thiemann2007}, where it is argued that one should refer to these inner products as inner products and not as transition amplitudes.}. Therefore, if one wants to insist that the quantities defined as transition amplitudes have some physical meaning and interpretation one cannot appeal to the notions and intuitions that applied to the case of the deparametrizable model. In this sense, the interpretation of these quantities as probabilities remains unclear, and the standard interpretations of quantum mechanics are not available in this case, as we lack the usual structure of a quantum state evolving in a Hilbert space\footnote{See \cite{Colosi2003}, where it is argued for such a probabilistic interpretation of `transition amplitudes' defined by means of the inner product of the physical Hilbert space. }. Moreover, in the double harmonic oscillator example it was clear that the inner product didn't have any straightforward interpretation as a transition amplitude and that it was signaling that the canonical quantization program simply failed to give a successful quantization of the theory. In the case of LQG this is a possibility one should also consider.

Finally, the problem of time also affects some cosmological models, which use symmetry reduced versions of either geometrodynamical or connection variables\footnote{I refer the reader to \cite{Ashtekar2006a} for a comparison of models using both kinds of variables.}. Given the simplicity of the models considered, it is common to see that the strategy followed for these models consists in deparametrizing the theory. For instance, it is common to see that the models describe the quantum evolution of some geometrical degrees of freedom in a time that is given by some scalar field\footnote{This kind of model can be found for instance in \cite{Ashtekar2006a} and \cite{Ashtekar2011}.}. The arguments above also work against this kind of model, and one should be cautious when inferring any physical prediction from them. 

Let me conclude this section by mentioning three other different attitudes that one may have towards the problem of time. First, Barbour \cite{Barbour1994,Barbour1994a,Barbour1999} endorses the view that the problem of time leads to a timeless ontology where there is no flow of time but just instants with some probability to happen. Second, a Bohmian can accept a static wavefunction, as long as they are able to define a guidance equation which describes a temporal evolution for some basic ontology of the quantum gravity theory\footnote{A sketch of how this would work for the case of LQG can be found in \cite{Vassallo2014}.}. Finally, we find proposals like \cite{Gryb2014,Gryb2016}, in which instead of quantizing general relativity one quantizes shape dynamics, which is an empirically equivalent theory but with a different set of symmetries. The hope is that as this theory doesn't show the same temporal reparametrization invariance as general relativity one can avoid the problem of time. 

\section{Conclusions} \label{conclusions}

In this paper I have argued that the problem of time affects all reparametrization invariant theories, but that the potential resolutions of it may work only for a subset of them, the deparametrizable ones. I have argued for this by comparing two examples: a deparametrizable model of a non-relativistic particle and the Jacobi action describing a system of two harmonic oscillators. This comparison showed that in the case of the double harmonic oscillator the most reasonable conclusion is that the canonical quantization of the model doesn't lead to a physically meaningful quantum theory.

For the case of general relativity I have argued that it is not a deparametrizable theory and that its configuration space doesn't seem to encode information about time in any way, just as happened for the case of the double harmonic oscillator. Therefore, I have argued that the theories of quantum gravity based on the resolutions of the problem of time that worked for deparametrizable models may be ill-founded.

\section*{Acknowledgments}
I want to thank the philosophy of physics reading group at UCSD, the Proteus group, Carl Hoefer, Brian Pitts and Daniele Oriti for their comments and discussions. This research is part of the Proteus project that has received funding from the European Research Council (ERC) under the  Horizon 2020 research and innovation programme (Grant agreement No. 758145) and of the project CHRONOS (PID2019-108762GB-I00) of the Spanish Ministry of Science and Innovation.


\bibliographystyle{acm}
\bibliography{sn-bibliography}


\end{document}